\documentclass[11pt]{article}

\usepackage{amsmath,amssymb}
\usepackage{graphicx}
\usepackage{epstopdf}
\newcommand{\be}{\begin{equation}}
\newcommand{\ee}{\end{equation}}
\newcommand{\bea}{\begin{eqnarray}}
\newcommand{\eea}{\end{eqnarray}}
\newcommand{\ba}{\begin{array}}
\newcommand{\ea}{\end{array}}

\newcommand{\eq}[1]{(\ref{#1})}

\begin{document}
%%%%%%%%%%%%%%%%%%%%%%%%%%%%%%%%

%%%%%%%%%%%%%%%%%%%%%%%%%%
\begin{titlepage}
\begin{flushright}
\end{flushright}

\vfill

\begin{center}
{\large \bf
Non-relativistic Holography and Singular Black Hole}
\vfill
{
Feng-Li Lin$^a$\footnote{\tt linfengli@phy.ntnu.edu.tw}
}
 and
{
Shang-Yu Wu$^b$\footnote{\tt loganwu@gmail.com}
}

\bigskip
{\it
$^a$Department of Physics,
National Taiwan Normal University,
Taipei, 116, Taiwan\\
}

and

{\it
$^b$ Department of Physics,
National Taiwan University,
Taipei, 106, Taiwan\\
}

\end{center}

\vfill
\begin{abstract}
 We provide a framework for non-relativistic holography so that a covariant action principle ensuring the Galilean symmetry for dual conformal field theory is given. This framework is based on the Bargmann lift of the Newton-Cartan gravity to the one-dimensional higher Einstein gravity, or reversely, the null-like Kaluza-Klein reduction. We reproduce the previous zero temperature results, and our framework provides a natural explanation about why the holography is co-dimension 2. We then construct the black hole solution dual to the thermal CFT, and find the horizon is curvature singular. However, we are able to derive the sensible thermodynamics for the dual non-relativistic CFT with correct thermodynamical relations. Besides, our construction admits a null Killing vector in the bulk such that the Galilean symmetry is preserved under the holographic RG flow. Finally, we evaluate the viscosity and find it zero if we neglect the back reaction of the singular horizon, otherwise, it could be nonzero.
\end{abstract}
\vfill

\end{titlepage}

%%%%%%%% BODY %%%%%%%%%%%%%
\setcounter{footnote}{0}
%%%%%%%%%%%%%%%%%%%%%%%%%%%%%%%

%%%%%%%%%%%
\section{Introduction}
%%%%%%%%%%%

  AdS/CFT correspondence based on holographic principle \cite{Maldacena:1997re} has provided an important tool to study the strongly coupled systems. For example, an lower bound on viscosity to entropy ratio (called KSS bound) has been derived via dual black hole physics \cite{Policastro:2001yc,Kovtun:2004de}.  In fact, Nature is fond of non-relativistic systems as far as the strong correlation is concerned such as the ultra-cold atoms at Feshbach resonance or quantum Hall effect. Moreover, the new technology advance in nano-physics has realized more tunable coupling systems, such as the Kondo effect in  quantum dots, where notable quantum phase transition occurs. It is then more desirable if one can apply the holographic principle to more strongly coupled/correlated systems in the condensed matters physics.   Recently, two papers have appeared in attempt to study the  holographic dual of non-relativistic conformal field theory (CFT). \cite{Son:2008ye,Balasubramanian:2008dm}. There, they proposed a background metric with the Schr\"{o}dinger symmetry (Galilean plus conformal symmetry) as its isometry by introducing an additional null-like Killing direction. Therefore, the non-relativistic CFTs is co-dimension two dual to the bulk theory.

  On the other hand, the generalization of the proposal in \cite{Son:2008ye,Balasubramanian:2008dm} to finite temperature has also been tried and studied in \cite{Herzog:2008wg,Adams:2008wt,Maldacena:2008wh}. In these papers, the temperature for the boundary CFT is turned on by null Melvin twisting the Schwarzschild-AdS black hole, from which the thermodynamics of the dual CFT is derived.  They also calculate the viscosity-to-entropy-density ratio, and it satisfies the KSS bound as expected. However, since the null Melvin twist transformation only modifies the asymptotic symmetry, therefore in \cite{Herzog:2008wg, Adams:2008wt, Maldacena:2008wh} the background metric has preserved the null-like Killing direction only on the boundary at infinity but not inside the bulk. This implies that the Galilean symmetry of the dual CFT is only preserved at UV end. As the holographic RG flow runs down toward IR, it is not clear how the Galilean symmetry can be realized explicity since there is no null Killing vector in the bulk.

  In this paper we try to preserve the Galilean symmetry in the bulk, i.e., our background metric will allow a null-like Killing vector in the bulk, by considering a dual black hole with singular horizon. The singularity is due to a no-go theorem, firstly proposed in \cite{Liu:2003cta,Hubeny:2002pj}, saying that there is no regular horizon if the bulk admits a null-like Killing vector. This is the price to pay to preserve the Galilean symmetry for the dual CFT through the RG flow. However, adopting the standard method, we will show that the singular horizon still can yield finite temperature and correct thermodynamics for the co-dimension two dual non-relativistic CFT.

  In order to construct the black hole solution, we should have an action principle for the dual gravity so that the field equations can be derived and used to construct the black hole. To achieve it, we first provide a framework in understanding the non-relativistic holography. Our basic idea is that the non-relativistic QFT should be holographically dual to the Newton-Cartan gravity in one-dimensional higher since the dynamical symmetry of QFT matches with the isometry of the gravity as in the relativistic case \cite{Maldacena:1997re}. The Newton-Cartan gravity \cite{Kuchar:1980tw},  though formally covariant, is restricted by the Galilean symmetry to have the absolute time, and hence awkward to implement the holography. On the other hand, it is known that the Schr\"{o}dinger symmetry can be obtained from the Kaluza-Klein (KK) reduction along a null-like direction \cite{Henkel:2003pu,Son:2005rv,Hassaine:2007sv}.  Therefore, we can  lift the Newton-Cartan gravity dual theory along a null Killing direction to one-dimensional higher covariant Einstein gravity, and the corresponding geometry is associated with the Bargmann group \cite{Duval:1984cj,Duval:1990hj,Julia:1994bs}. We can also turn the other way around, given a holographic dual pair of relativistic theories, we can perform the null-like KK reduction to obtain the holographic dual pair of the non-relativistic theories.

  Though the singular horizon is usually controversial in semi-classical gravity, we try to argue that we can still yield sensible thermodynamics for the dual non-relativistic CFT if some appropriate conditions are assumed. We hope our results will serve as a toy example in exploiting the singular black hole for holography.

  Our paper is organized as follows. In next section we will delineate our framework for the non-relativistic holography based on the Bargmann lift and null-like KK reduction \cite{Duval:1984cj,Julia:1994bs}. An action principle and the field equation for the gravity are given in this framework. In section 3 we construct the black hole solution, and find the singular horizon. We then use the usual boundary stress tensor formalism to derive the thermodynamics of the dual CFT, which matches with the one for the non-relativistic thermal gas. In section 4 we try to evaluate the viscosity and find it zero using the conventional method of holographic hydrodynamics. We discuss the validity of our results. Finally, we conclude in section 5 along with some discussions about the singular horizon and viscosity.

%%%%%%%%%%%%%%%%%%%%%%%%%%%%%%%%%%%%%%%%%%%%%%%%%%%%%%%%%%%%%%%%%%%%%%%%%%%%
\section{A Framework}
%%%%%%%%%%%%%%%%%%%%%%%%%%%%%%%%%%%%%%%%%%%%%%%%%%%%%%%%%%%%%%%%%%%%%%%%%%%%

  Generalizing the spirit of AdS/CFT correspondence \cite{Maldacena:1997re} by identifying the isometry of the AdS space with the kinematic symmetry of the dual CFT, it is natural to speculate that the gravity dual of a co-dimension one non-relativistic CFT should be the Newton gravity so that they share the same Galilean symmetry. However, since the Newton gravity is not a covariant theory, it is not clear what is the geometric formulation of the background spacetime. This makes it ambiguous to identify the conformal isometry in the gravity side. Indeed, there is a covariant formulation of Newton gravity, known as the Newton-Cartan gravity \cite{Kuchar:1980tw}, in which the Newton-Poisson equation can be put into the covariant form by analogy with Einstein equation
  \be
  R^{(NC)}_{a b}=4\pi G_N \rho \psi_a \psi_b, \qquad a,b=0,\cdots,3,
  \ee
  where $\rho$ is the mass density and $\psi_a$ are the time-like vector charactering the absolute Newton's time. Despite of the covariant formulation, the metric of the background spacetime in Newton-Cartan gravity is degenerate with signature $(0,+,+,+)$ to have the Galilean symmetry manifested. This then implies the Ricci tensor $R^{(NC)}_{\mu\nu}$ is highly constrained so that one should impose non-trivial conditions among its components, and also on the Newton connection. These constraints make the Newton-Cartan formulation hard to implement in dealing practical problems, e.g., the action principle that Newton-Poisson equation can be derived is unclear.

  It is then desirable to have a covariant formulation of Einstein-type gravity for the dual non-relativistic CFT, moreover, with an action principle incorporated.

\subsection{Bargmann Lift of Newton-Cartan Gravity}

  In \cite{Duval:1984cj}, it was shown that four-dimensional  Newton-Cartan theory can be formulated on a five-dimensional space-time called Bargmann manifold. By such a lifting, one can overcome the shortcoming of lacking the true non-degenerate metric structure in Newton-Cartan theory. Moreover, an action principle can be possibly formulated.  In \cite{Duval:1984cj}, a Bargmann manifold $\widetilde{M}$ with metric $g_{\mu\nu}$ of Lorentzian signature $(-,+,+,+,+)$ is defined by lifiting a four-dimensional connected smooth manifold $M$ along a null-like direction generated by a covariant constant Killing vector $\xi^{\mu}$, i.e.,
\be
g_{\mu \nu}\,\xi^{\mu}\,\xi^{\nu}=0, \qquad \nabla_{(4+1)}^{\mu}\xi^{\nu}=0, \qquad \mu,\nu=0,\cdots,4.
\ee

  In practice, Bargmann lift is more than just a reformulation of Newton-Cartan gravity. As proposed in \cite{Duval:1984cj}, the field equation that gives the metric on Bargmann manifold can be obtained from the usual variation of the Einstein-Hilbert action with the slight modification:
\be
\label{actBarg}  \int_{\widetilde{M}}\,d^5x \,\sqrt{-g}\, \big( R -\lambda\,g_{\mu\nu}\xi^{\mu}\xi^{\nu} \big)\,\,.
\ee
  The introduction of the additional term with $\lambda$ as a Lagrange multiplier is to impose a constraint \footnote{It seems that we should also add the Lagrange multiplier term for the covariant constancy of $\xi^{\mu}$. However, we can save it by directly requiring the metric is independent of the null-like coordinate generated by $\xi^{\mu}$.}, which ensures that the null-likeness of the chosen $\xi^{\mu}$ will be preserved by the Bargmann metric solving the field equation derived from \eq{actBarg}
\be
 R^{\mu\nu}- \frac{1}{2} R  g^{\mu\nu}- \lambda \xi^{\mu}\xi^{\nu} = 0.
\ee
  After taking the trace, we find the scalar curvature is zero. Therefore, the final field equation is in the suggestive form
\be
\label{feqbarg} R_{\mu\nu} = \lambda\, \xi_{\mu} \xi_{\nu}\,\,,
\ee
  where $\xi_{\mu}=g_{\mu\nu} \xi^{\mu}$ is time-like. This looks like the Newton-Poisson equation of Newton-Cartan gravity if we identify the Lagrangian multiplier function $\lambda$ as the mass density
\be
\lambda \equiv 4\pi G_N \rho\,\,.
\ee
  Moreover, the Ricci tensor $R_{\mu\nu}$ is strictly equal to $R_{ab}$ after null-like dimensional reduction along $\xi^{\mu}$ \cite{Julia:1994bs} with some proper gauge-fixing choice of the Newton connection, see \cite{Duval:1984cj} for details.

  An illustrative example of the Bargmann metric is
\be
\label{Umetric} ds^2 = -2 U(t,x^i) dt^2 - 2 dt dv + dx^i dx^i\,\,, ~~\\~~ i=1,2,3,
\ee
  where $v$-direction is the null-like KK direction by choosing $\xi^{\mu}=\delta^{\mu,v}$. Note that, this is a pp-wave metric, and $U(x^i,t)$ is nothing but the Newton's potential.
  By plugging this metric into the equation \eq{feqbarg}, one finds
\be \nabla^2\,U(t,x^i) = 4\pi G_N \rho(t,x^i)\,\,, \ee
  which is exactly the Newton-Poisson equation. A trivial example is $\lambda=0$, then $U=1$ and the resulting Bargmann metric is flat.

%%%%%%%%%%%%%%%%%%%%%%%%%%%%%%%%%%%%%%%%%%%%
\subsection{Conformal Symmetry}
%%%%%%%%%%%%%%%%%%%%%%%%%%%%%%%%%%%%%%%%%%%%

  In \cite{Duval:1990hj}, it was shown that the conformal transformations ${\cal C}$ of the 5-dimensional flat Bargmann metric that commute with the structure group (i.e., ${\cal C}\xi=\xi$) form a 13-dimensional Lie group, the so-called extended Schr\"{o}dinger group. Furthermore, it is clear that by introducing $U(t,x^i)$, some (or all) of symmetries will be broken. However, it was  also shown in \cite{Duval:1990hj} that if $U$ is the inverse square potential i.e., $U \sim 1/|\vec{x}|^2$, the following SL(2,R) subgroup of Sch(3) is preserved:
\bea
D:& (t, x^i)\, ~\rightarrow~& (s^2 t, s x^i)\,\,,  \\
C:& (t, x^i)\, ~\rightarrow~& (\frac{t}{1+\lambda t}, \frac{x^i}{1+\lambda t})\,\,, \\
H:& (t, x^i)\, ~\rightarrow~& (t+e, x^i)\,.
\eea
  Later on, the similar conformal symmetry can be exploited for the non-relativistic holography.

%%%%%%%%%%%%%%%%%%%%%%%%%%%%%%%%%%%%%%%%%%%%
\subsection{Non-relativistic Holography}
%%%%%%%%%%%%%%%%%%%%%%%%%%%%%%%%%%%%%%%%%%%%

  As long as the field theory dynamics is concerned, one can also perform the Bargmann lift to have relativistic formulation of the conformal dynamics, for examples and general discussions, see \cite{Henkel:2003pu, Son:2005rv, Hassaine:2007sv, Duval:1990hj, Leiva:2003kd}. Indeed, this is what was used in \cite{Son:2008ye,Balasubramanian:2008dm} to evaluate the GKP/W-type correlators \cite{Gubser:1998bc,Witten:1998qj} for a non-relativistic CFT from its bulk holographic co-dimension two scalar field theory. Their results reproduce the correct scaling behavior of a non-relativistic CFT.

\begin{figure}[ht]
\begin{center}
\includegraphics[width=15cm]{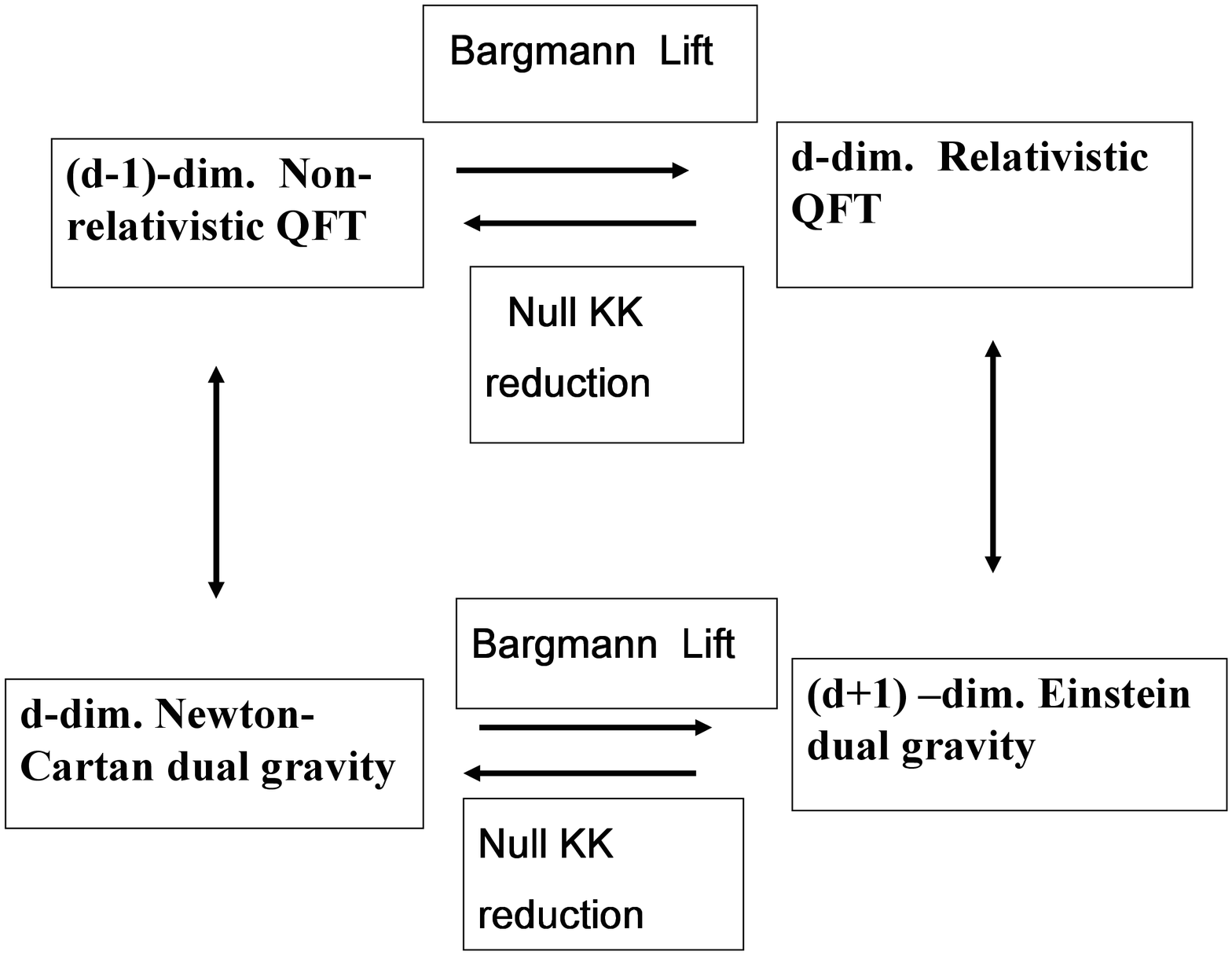}
\caption{The framework of non-relativistic holography}
\label{fig1}
\end{center}
\end{figure}

  The previous discussions on the Bargmann lift suggest a possible covariant formulation of Einstein-type gravity dual for a non-relativistic CFT. As conjectured, the Newton-Cartan gravity with a negative cosmological constant should be dual to a co-dimension one non-relativistic CFT. However, the Newton-Cartan formulation is hard to implement in practice, so we can lift it to the Bargmann  formulation, in which an action principle is possible. Conversely, we can reduce from the Bargmann formulation to the Newton-Cartan one  by performing the null-like KK reduction, see \cite{Julia:1994bs} for detailed discussions. The introduction of the Bargmann lift results in a co-dimension two holography. This framework for the non-relativistic holography can be summarized  in Fig. \ref{fig1}.

  In order to utilize the well-studied AdS holography for our framework, we introduce a negative cosmological constant in action \eq{actBarg}, by which we hope it can describe a co-dimension two dual non-relativistic CFT. The corresponding field equation becomes:
\be\label{ABfeq}
R_{\mu\nu}- \frac{1}{2}R g_{\mu\nu} + \Lambda g_{\mu\nu}- \lambda \xi_{\mu}\xi_{\nu}=0\,.
\ee
  Similarly, once a null-like Killing vector is chosen, we can reduce this equation to the Newton-Poisson equation with a negative cosmological constant by properly fixing the gauge of the Newton connection. It is easy to see that the constraints on the Newton-Cartan Ricci tensor take the same form as in the flat space. Therefore, the Bargmann formulation with a cosmological constant should be the same as the one without cosmological constant, for example, see the recent discussion in \cite{Duval:2008jg}. However, the null-like KK reduction is quite non-trivial even for the asymptotically flat space due to the degenerate metric. Especially, to reconcile the diffeomorphism and local Galilean symmerty, some special care should be taken in choosing the frame and gauge fixing the connection as discussed in \cite{Julia:1994bs}. Therefore, a more complete treatment for the null-like KK reduction in the context of non-relativistic holography should be further explored.   Here, we will just take a more pragmatic approach and leave the formal issue aside.

  Once the null-like Killing vector $\xi^{\mu}=\delta^{\mu,v}$ is chosen, we can obtain the generalized Bargmann metric for a given $\lambda$.  The most trivial solution is for $\lambda = 0$ and it can be written as:
\be ds^2 = \frac{1}{z^2} (-2 dt dv + dx_{d_s}^2+ dz^2).
\ee
  This is, as expected, just the AdS space in the Poincar\'{e} coordinate, which has the maximal isometry as the extended Schr\"{o}dinger group for the flat Bargmann metric.

  Similarly to what we have done for the $\Lambda=0$ case, one can introduce the ``Newton potential" in the metric
\be\label{adsU}
ds^2 ~=~ \frac{1}{z^2}\,( -2\,U(t,x^i,z)\,dt^2 - 2 dt dv + dx_{d_s}^2 + dz^2 ),
\ee
  it is easy to see that this metric is indeed a solution of equation \eq{ABfeq} if $U(t,x^i,z)$ satisfies:
\be
\nabla^2\,U(t,x^i,z) =-\lambda(t,x^i,z),
\ee
  where the Laplacian is defined with respect to the metric \eq{adsU}.

  As for the $\Lambda=0$ case, we may ask which $U(t,x^i,z)$ can preserve the extended Schr\"{o}dinger group so that it is consistent with the procedure of embedding $Sch(d_s)$ (Schr\"{o}dinger group in $d_s$ spatial dimensions) in $O(d_s+2,2)$ in \cite{Son:2008ye}. In this case, the condition of preserving the bundle structure becomes the selection of subgroup formed by the generators that commute with the light-cone momentum associated with $\xi^{\mu}$. This is just the non-relativistic analogy of embedding the conformal group into the isometry of AdS space.  Since the metric \eq{adsU} is simply Weyl-related to \eq{Umetric}, following the lesson for the $\Lambda=0$ case we should consider that $U$ is independent of $x^i$ but still an inverse square potential:
\be
U =\frac{1}{z^2}.
\ee
  This corresponds to choose $\lambda=d(d-1)/2$ where $d=d_s+2$. Therefore, the metric that exhibits a full Schr\"{o}dinger group is the following:
\be
ds^2 =  \frac{1}{z^2}\,( -\frac{1}{z^2} dt^2 - 2 dt dv + dx^2_{d_s} + dz^2 ).
\ee
  See also \cite{Duval:2008jg} for the related discussions. Note that, this is the same metric that obtained in \cite{Son:2008ye, Balasubramanian:2008dm} for the non-relativistic holography of the cold atoms in $d_s+1$ dimensions, there the GKP/W-type correlators are also constructed. Also this metric belongs to the class of Siklos space-time,
\be
ds^{2}=\frac{1}{r^{2}}(-F(\vec{x},r,t)dt^{2}+2dtdv+d\vec{x}^{2}+dr^{2}),
\ee
which can be interpreted as the gravitational wave traveling along AdS. See \cite{Duval:2008jg} for related discussions.

%%%%%%%%%%%%%%%%%%%%%%%%%%%%%%%%%%%%%%%%%%%%
\section{Singular Black Hole}
%%%%%%%%%%%%%%%%%%%%%%%%%%%%%%%%%%%%%%%%%%%%

   Based on the action principle of the Bargmann lift with a negative cosmological constant, we would like to find the black hole solution corresponding to turning on the temperature of the dual CFT and the related thermodynamical quantities. However, the task is obscured by a no-go theorem proposed in \cite{Liu:2003cta,Hubeny:2002pj}: there will be no regular black hole horizon if the spacetime admits a null-like Killing vector. The original argument in \cite{Liu:2003cta,Hubeny:2002pj} is for the asymptotically flat space, namely, the pp-wave, however, it seems that the no-go theorem still holds with a negative cosmological constant included.  Unlike the approach taken in \cite{Herzog:2008wg,Adams:2008wt,Maldacena:2008wh} in which they gave up the null-like Killing vector in the bulk, we will endeavor to construct the singular black hole in our Bargmann lift framework, and derive from it the thermodynamics of the dual CFT. An immediate advantage over the approach in \cite{Herzog:2008wg,Adams:2008wt,Maldacena:2008wh}, our solution preserves the null-like Killing vector in the bulk which implies that the Galilean symmetry is preserved under the holographic RG flow.  Of course, we will face many tricky issues regarding the singular black hole.

\subsection{Finding the solution}

  Our starting point for the black hole solution is the following action of $(d+1)$-dimensional Anti-de Sitter (AdS) gravity with a scalar
\be
\int d^{d+1}x \sqrt{-g} [R-{1\over 2} (\partial \phi)^2- {1\over 2}d(d-1) - \lambda \xi^{\mu}\xi_{\mu}],
\ee
and the resultant equations of motion are \footnote{We set the AdS curvature radius $R_{ads}$ to one for the moment, and will recover it later on when
it is necessary.}
\be
R_{\mu\nu}+d g_{\mu\nu}-\lambda \xi_{\mu} \xi_{\nu}={1\over 2} \partial_{\mu}\phi \partial_{\nu}\phi
\ee
and
\be
\partial_{\mu}(\sqrt{-g} \partial^{\mu} \phi)=0.
\ee
Here the Largrangian multiplier term is added to impose the condition for the null-like KK reduction. We choose the
null-like Killing vector to be $\xi^{\mu}=\delta^{\mu,v}$ for the metric ansatz
\be
ds^2=e^{2A(r)}\left(-2dt dv+ {\cal H}(r) dt^2\right)+e^{2B(r)}dr^2+e^{2C(r)} dx^2_{d-2}.
\ee
The equations of motion then reduce to
\be
e^g \phi'=const.,
\ee
and
\be
A''+A'g'=d e^{2B},
\ee
\be
C''+C'g'=d e^{2B},
\ee
\be
A''+A'^2-A'B'+{d-2\over 2}(C''+C'^2-C'B')={d\over 2}e^{2B}-{1\over 4}\phi'^2,
\ee
\be\label{Heom}
{\cal H}''+{\cal H}' g'+2{\cal H}(A''+A'g')-2d{\cal H} e^{2B}+2\lambda e^{2A+2B}=0,
\ee
and
\be
g=2A-B+(d-2)C.
\ee
In the above $':={d\over dr}$.

Taking the ansatz
\be
A(r)=a\ln k(r)+ \alpha \ln r, \qquad B(r)=b \ln k(r) +\beta \ln r, \qquad C(r)=c \ln r,
\ee
and then solve for $a,b,c,\alpha,\beta$ and $k(r)$ from the above equations.

 It is straightforward to find the nontrivial solutions with the above ansatz. After some algebra, the solution we find is
\be\label{bhsol}
ds^2=r^2\sqrt{f(r)}(-2dt dv+{\cal H}(r) dt^2)+{dr^2\over r^2 f(r)}+r^2 dx^2_{d-2}
\ee
and
\be
e^{2\phi(r)}=f(r),
\ee
where
\be
f(r)=1-{r^d_H\over r^d}
\ee
and
\be
\lambda(r)=-{1\over 2r^{1+d}\sqrt{f}}({\cal H}'r^{1+d} f)'\label{lambda}.
\ee
In the above the radial coordinate is in the unit of AdS curvature radius $R_{ads}$. Note that, $\phi\rightarrow 0$ as $r\rightarrow \infty$ so it yields no contribution to boundary action in the holographic consideration.

  Our solution has some interesting features as follows:

(1) As in \cite{Herzog:2008wg,Adams:2008wt,Maldacena:2008wh}, we also need to turn on the dilaton profile to obtain the
above quasi black hole solution. This implies that the temperature effect will break the conformal symmetry and induce nontrivial
RG flow.

(2) Unlike the solution considered in \cite{Herzog:2008wg,Adams:2008wt,Maldacena:2008wh}, our solution preserves
the null-like Killing direction even in the bulk not only being restricted to the asymptotic boundary infinity. This implies that
as we change the RG scale from UV to IR, we can still have the dual QFT with Galilean symmetry.

(3) Our black hole is quasi since the horizon can be reached only in the radial direction, not in the $t$ and $v$ directions. This is consistent
with the fact that there is no notion of black hole in the Newtonian gravity obtained from the null-like KK reduction.
This can be seen from the criterion discussed in \cite{Liu:2003cta}, see also \cite{Hubeny:2002pj}. In \cite{Liu:2003cta}, they assume that near
the horizon, we have
\be
e^{2A}\sim (r-r_H)^{2a}, \qquad e^{2B}\sim (r-r_H)^{2b}, \qquad  {\cal H}\sim (r-r_H)^{2h} sign({\cal H})
\ee
For our solution, we have \footnote{The exponent $h=1/4$ will be determined in the next subsection.}
\be
a={1\over 4}, \qquad b=-{1\over 2}, \qquad  h={1\over 4}, \qquad sign({\cal H})=-1.
\ee
As shown in \cite{Hubeny:2002pj}, for the coordinate time of a static observer to be finite in reaching the horizon along the radial direction we need
\be
a+b>-1, \qquad \mbox{if}\;\; h\ge 0 \;\; \mbox{or}\;\; {\cal H}\sim \log(r-r_H),
\ee
and
\be
a+b+|h|>-1, \qquad \mbox{if}\;\; h< 0
\ee
This is indeed the case for our solution.

 On the other hand,   for the coordinate time to be finite in reaching the horizon along the $t$ and $v$ directions we need
\be
a-b>1, \qquad \mbox{if}\;\; h\ge 0 \;\; \mbox{or}\;\; {\cal H}\sim \log(r-r_H),
\ee
and
\be
a-b>1+|h|, \qquad \mbox{if}\;\; h< 0
\ee
which is not obeyed by our solution.

   In summary, a static observer can ``see" the horizon only along the radial direction. This, however, will holographically provide the finite temperature environment for the dual CFT, and we will justify this later on by working out a sensible thermodynamics for the dual CFT.

(4) Moreover, the curvature invariants $R_{\mu\nu\lambda\sigma}R^{\mu\nu\lambda\sigma}$, $R_{\mu\nu}R^{\mu\nu}$ and $R^2$ are singular as $(r-r_H)^{-2}$ at $r=r_H$, so $r=r_H$ is a naked singularity.  At first sight, it is not clear if there is sensible thermodynamics associated with the singular horizon. However, we will see that we can indeed obtain finite thermodynamical quantities by using the conventional boundary stress tensor method.

\subsection{Thermodynamics}

   Now we will extract the thermodynamics out of the metric. Since the black hole horizon is singular, it is not clear if we can have sensible thermodynamics associated with or not. The first thing to check is to evaluate the temperature. One naive way to obtain the temperature is to require the absence of the conical deficit for the following general black hole metric
\be\label{generalmetric}
ds^2=F(r)d\tau^2+{dr^2\over G(r)}+ \cdots
\ee
where $\cdots$ denote the irrelevant terms for temperature extraction. Near the horizon we can write the above into a disk
without conical deficit by requiring the black hole temperature $T$ to be
\be
T={d F\over dr}\sqrt{G(r) \over F(r)}\mid_{r=r_H}.
\ee
Therefore, to have a well-defined $T$ we should require $G/F$ to be finite at $r=r_H$.

   One subtlety for our black hole metric \eq{bhsol} is the non-vanishing $g_{tv}$ which is absent in \eq{generalmetric}.  However, since we will KK reduce along the $v$-direction for the holographic duality, we can simply drop this term. Then, in our case we have $G(r)=r^2f(r)$ and $F(r)=r^2\sqrt{f(r)}|{\cal H}(r)|$, and the latter will depend on $\lambda(r)$.  For the case with $\lambda=-d(d-1)/2$ one can show that ${\cal H}(r_{H})$ will blow up, and we will not get a sensible $T$. Instead,  we should require ${\cal H}(r)=-C_t^2 r^{q-2}\sqrt{f(r)}$ for some constant $p$ and $C_t$, which can be obtained by tuning $\lambda(r)$ through \eq{lambda}. Then we find
\be
T={C_t d\over 2\pi} {{\bar r}_H^{q\over 2}\over R_{ads}}\label{BH T}
\ee
where we have introduced the normalized horizon radius ${\bar r}_H=r_H/R_{ads}$.

 For the moment, it seems that we have arbitrariness in getting different temperature by tuning $\lambda(r)$. We will see that this is not case if we require the first law of the black hole thermodynamics holds.

 One may worry about the above procedure in obtaining $T$ by simply dropping off the $g_{tv}$ term. To justify the answer and see the subtlety, we evaluate the surface gravity of the horizon. To do this, we need to choose a time-like Killing vector $k^{\mu}$ which becomes null-like on the horizon, and the surface gravity $\kappa$ is given by
\be
k^{\mu}\nabla_{\mu} k^{\nu}|_{r=r_H}=\kappa k^{\nu}|_{r=r_H}.
\ee
One can then follow the standard procedure by evaluating the above relation in the Finkelstein-Eddington coordinate to yield $\kappa$. If we choose $k^{\mu}=\delta^{\mu,t}$, we will get $\kappa$ which agrees with \eq{BH T} from the conical method. However, one subtlety arises \footnote{We thank Chi-Wei Wang for pointing this out.}: Since we have a null-like Killing vector $\xi^{\mu}=\delta^{\mu,v}$, any normalized time-like vector $a k^{\mu} +b \xi^{\mu}$ also becomes null on the horizon. It seems that the surface gravity is not unique. Moreover, it is easy to see that all the above time-like Killing vectors except $k^{\mu}$ will yield infinite surface gravity. Therefore, it seems ambiguous in defining the surface gravity and the black hole temperature. In any case, we are dealing with a singular black hole but aiming for a holographic interpretation of the boundary non-relativistic CFT. Instead of resolving the ambiguity definitely, we will follow the guidance of the holographic principle by noting that the null-like Killing direction is just an auxiliary to yield Galilean symmetry of the dual CFT via KK reduction. Therefore, we should keep this direction intact while considering the thermodynamics for the dual theory and $k^{\mu}$ should be the choice for such a purpose. In sum, though there is an ambiguity in defining the surface gravity for the black hole horizon, it is unambiguous to define a finite temperature for the dual CFT via a special choice of time-like Killing vector to evaluate surface gravity.  This feature is peculiar for our framework of non-relativistic holography via singular black hole.

   Given a temperature, we will obtain the thermodynamical quantities. First, the entropy is given by the Bekenstein-Hawking entropy \footnote{It is not clear if a singular horizon still obeys the area law or not, here we simply assume it is the case. With hindsight, if the singularity can be resolved and a stretched horizon is formed, then the area law should hold as long as the horizon size is far larger than the resolution scale.} so the entropy density of the dual CFT is
\be
{s\over L_v}={{\bar r}_H^{d-2}\over 4 G_{d+1}} ={1\over 4G_{d+1}} ({2\pi T R_{ads}\over C_t d})^{2(d-2)\over q}.
\ee
where $L_v$ is the size of the $v$-direction.

 Next, we can use the formalism of the boundary stress tensor \cite{Balasubramanian:1999re} to evaluate the energy and pressure density of the dual CFT.

The boundary stress tensor is given by $\Theta_{\mu\nu}=-(\nabla_{\mu}\hat{n}_{\nu}+\nabla_{\nu}\hat{n}_{\mu})$, where $\hat{n}_{\mu}$ is an inward unit vector normal to the boundary, i.e., $\hat{n}_{\mu}=-\sqrt{g_{rr}}\delta_{\mu,r}$. It is then straightforward to calculate the non-vanishing components of the boundary stress tensor, which result in the expectation value of stress tensor of the dual CFT
\begin{equation}
\langle(T_{bdy})^{\mu}_{\nu}\rangle={1\over 8\pi G_{d+1}} \sqrt{-\gamma}(\Theta^{\mu}_{\nu}-\delta^{\mu}_{\nu}\Theta)+ \mbox{covariant counter terms},
\end{equation}
where $\gamma$ is the determinant of the boundary metric.  The covariant counter terms play the role of the UV cutoff to yield finite stress tensor.

Consider the metric of our singular black hole
\begin{equation}\label{finalhole}
ds^{2}=-C_t^2 {\bar r}^{q}fdt^{2}-2{\bar r}^{2}\sqrt{f}dtdv+\frac{d{\bar r}^{2}}{{\bar r}^{2}f}+{\bar r}^{2}dx_{d-2}^2
\end{equation}
where ${\bar r}=r/R_{ads}$ is the normalized radial coordinate. Evaluating the boundary stress tensor of this metric and plus appropriate covariant counter terms
\be
-{1\over 8\pi G_{d+1}}(2d-2) \sqrt{-\gamma}\delta^{\mu}_{\nu}+c_0 \xi^{\mu}\xi_{\nu},
\ee
we obtain the energy density and pressure of the dual CFT in thermal equilibrium at temperature \eq{BH T} as follows
\be
{\rho\over L_v}:=-T^t_t={1\over 8\pi G_{d+1}}({d\over 2}-1){{\bar r}^d_H \over R_{ads}}, \qquad {p\over L_v}=:T^{x_i}_{x_i}={1\over 8\pi G_{d+1}}{{\bar r}^d_H \over R_{ads}}.
\ee
Note that relation $\rho=({d\over 2}-1)p$ could be implied by the non-relativistic conformal invariance \footnote{At least, for d=4, we have $\rho=p$ as expected in \cite{Herzog:2008wg,Adams:2008wt}.}.
Furthermore, we should require the first law of thermodynamics $\rho+p=Ts$ to hold, this yields
\be\label{firstlaw1}
q=4, \qquad C_t={1\over 2}.
\ee
This gives the correct scaling behavior for the non-relativistic thermal gas in $d_s=d-2$ spatial dimensions, i.e.,
\be
\rho \sim T^{{d_s\over 2}+1}, \qquad s \sim T^{d_s\over 2}.
\ee

  Few remarks are in order about our results:

 (1) Though it seems the starting backgrounds for the dual CFT have some arbitrariness in choosing $\lambda(r)$, we find that the arbitrariness is lifted by the consistency of thermodynamics required by the first law. From \eq{lambda} and ${\cal H}=-r^2\sqrt{f}/4$ implied by \eq{firstlaw1}, one can determine $\lambda(r)$.

 (2) Our profile of $\lambda(r)$ and $\phi(r)$ should be some matter sources of the dual Newton-Cartan gravity after KK reduction along null-like direction $v$. Our results suggest that only some particular matter sources in Newton-Cartan gravity can yield consistent thermodynamics for the dual CFT.

 (3) Beside the $T^t_t$ and $T^{x_i}_{x_i}$, $T^v_t$ is also non-vanishing but cannot be regularized by the covariant UV counter terms. However, unlike $T^t_v$ which corresponds to the particle number density in dual CFT, there is no proper interpretation for $T^v_t$ in dual CFT so that its UV divergence may not be physically relevant. Instead we have $T^t_v=0$ due to the absence of the KK particle condensation. It would be interesting to see how to achieve the vacuum of non-zero particle number density by turning on some bulk configuration, such as the one in Sakai-Sugimoto model \cite{Lin:2008vv}.

\section{A trial for Viscosity}

  Though our black hole has singular horizon, we have found a well-defined the thermodynamics for the dual CFT. The next step is to see if one can obtain some transport coefficients of the hydrodynamics such as the shear viscosity of the dual CFT. For case with regular black hole in dual gravity, there are many ways to derive the viscosity, and they all saturate the KSS bound for many different holographic dual quantum field theory. The deep reason for such a universality is deeply related to universality of the low energy absorption cross section of the bulk black hole \cite{Das:1996we} as pointed out in \cite{Kovtun:2004de}.  However, for the singular black hole, the above conclusion may break down because the perturbations around the horizon to obtain the viscosity may cause large back-reactions due to the strong curvature. Also, the recent study \cite{Bhattacharyya:2008xc} in holographically deriving the bulk geometry from the hydrodynamics of the boundary CFT suggests the horizon should be regular under the derivative expansions. Despite of these seemingly obstacles, we instead view our singular black hole as a test ground for the KSS bound.  In this section, we will just ignore the issue of back-reaction and pretend the hydrodynamical approximation remains good for singular horizon, and then evaluate the viscosity by following the methods in \cite{Policastro:2001yc,Kovtun:2004de,hydrodynamics}.

 For simplicity, we will consider $d=4$ case, the generalization to other dimensions should be straightforward.
 The shear viscosity can be obtained via the Kubo's formula
\begin{equation}
\eta=\lim_{\omega\rightarrow0}ImG^{R}_{12,12}(\omega,\vec{k}=0),
\end{equation}
where $G^{R}_{12,12}$ is the retarded Green's function
\begin{equation}\label{green1}
G^{R}_{12,12}(\omega, \vec{k}=0)=-i\int d^{4}xdt{e^{i\omega t}\theta(t)\langle T_{12}(t,\vec{x})T_{12}(0,0)\rangle}.
\end{equation}
As proposed in \cite{Policastro:2001yc,Kovtun:2004de}, the two point function can be obtained by the method of GKP/W for the linear perturbation of the bulk metric, $\Phi\equiv h^{2}_{1}$. It turns out the linearized Einstein equation for $\Phi$ is just the scalar field equation. In \cite{Kovtun:2004de}, it was also pointed out that the above method leads to
\be
\eta={\sigma_{abs}|_{\omega\rightarrow 0}\over 16\pi G_N}
\ee
where $\sigma_{abs}|_{\omega\rightarrow 0}$ is the low energy absorption cross section of the bulk black hole, which is determined by the ratio between fluxes at horizon and spatial infinity. Therefore, we should solve the scalar field equation on the black hole background and evaluate the fluxes.

  For our background metric \eq{finalhole} and \eq{firstlaw1} with $d=4$, after setting $\Phi=e^{-i(\omega t-\ell v)}\phi_\ell(r)$ where $\ell$ is the null-like KK momentum, the scalar field equation can be brought into the form of Schr\"{o}dinger equation (For the time being, we set $R_{ads}=1$.)
\be
\partial_{\rho}^2 \phi_{\ell} -[{\ell^2 r^8_H \over 4} {f\over (1-f)^2} +2\ell \omega r^6_H {\sqrt{f}\over (1-f)^{3/2}}] \phi_{\ell}=0
\ee
where
\be
-\infty < \rho:={1\over 4r^4_H} \ln f <0,\qquad f=1-{r^4_H\over r^4}.
\ee

  In the near horizon region, i.e., $\rho\rightarrow -\infty$, $\phi_{\ell}$ approaches a constant if we require it is regular at the horizon. On the other hand, in the far region, i.e., $\rho \rightarrow 0$, the asymptotic behavior of $\phi_{\ell}$ will depend on the sign of $\ell$. If $\ell>0$, $\phi_{\ell} \rightarrow C_1 \sqrt{-\rho}I_{\sqrt{4+\ell^2/4}}(2\sqrt{\ell\omega}(-\rho)^{-1/4}) +C_2 \sqrt{-\rho}K_{\sqrt{4+\ell^2/4}}(2\sqrt{\ell\omega}(-\rho)^{-1/4})$ which are real and non-oscillatory, where $I_{a}(x), K_{a}(x)$ are modified Bessel functions. Matching with the near horizon region to determine the integration constants $C_{1}$ and $C_{2}$, it implies that the solution is real and there is no flux.  This leads to zero $\sigma_{abs}|_{\omega\rightarrow 0}$, and thus vanishing viscosity.   On the other hand, for $\ell<0$, $\phi_{\ell} \rightarrow C_1 \sqrt{-\rho}J_{\sqrt{4+\ell^2/4}}(2\sqrt{\ell\omega}(-\rho)^{-1/4}) +C_2 \sqrt{-\rho}Y_{\sqrt{4+\ell^2/4}}(2\sqrt{\ell\omega}(-\rho)^{-1/4})$, where $J_{a}(x), Y_{a}(x)$ are the first kind and second kind of Bessel function respectively, can be imaginary and oscillatory. However, it cannot match with the constant behavior at the near horizon region. This implies no consistent solution, and thus no absorption and zero viscosity.

   Our result seems violating the KSS bound on $\eta/s$. It deserves to use the other way to evaluate it.  As known, one can also compute the shear viscosity from the hydrodynamics \cite{hydrodynamics} by evaluating the two point function in \eq{green1} $\grave{a}$ la GKP/W. It is convenient to write the metric in the following form
\begin{equation}
ds^{2}=-\frac{(\pi TR_{ads})^{2}}{4u^{2}}f(u)dt^{2}-\frac{2\pi TR_{ads}}{u}\sqrt{f(u)}dtdv+\frac{\pi TR_{ads}}{u}(dx^{2}+dy^{2})+\frac{R^{2}_{ads}}{4u^{2}f(u)}du^{2},
\end{equation}
where $u:=r^2_H/r^2$ so $f(u)=1-u^{2}$.

The scalar field equation becomes
\begin{equation}
\phi''_{\ell}(u)-\frac{1+u^{2}}{uf}\phi'_{\ell}(u)+(\frac{\omega \ell R_{ads}}{4\pi Tuf^{3/2}}-\frac{\ell^2 R_{ads}^{2}}{16u^{2}f})\phi_{\ell}(u)=0 \label{wave equation}
\end{equation}
In the new coordinate, the boundary is at $u=0$, and the horizon is at $u=1$. According to the recipe in \cite{hydrodynamics}, we assume $\phi_{\ell}(u)=(1-u)^{\delta}F_{\ell}(u)$, and we find the wave equation is
\begin{equation}
F''_{\ell}(u)-(\frac{1+u^{2}}{uf}+\frac{2\delta}{1-u})F'_{\ell}(u)+(\frac{\delta(\delta-1)}{(1-u^{2})^{2}}+\frac{\delta(1+u^{2})}{uf(1-u)}+\frac{\omega \ell R_{ads}}{4\pi Tuf^{3/2}}-\frac{\ell^2 R^{2}_{ads}}{16u^{2}f})F_{\ell}(u)=0.
\end{equation}
To determine $\delta$, it requires the most singular term vanishes at $u=1$, such that we find $\delta=0$. This means there is no phase factor, and the wave function becomes constant at the horizon. In fact, this already suggests there is no incoming flux, and the absorption cross section is zero, i.e.,  zero shear viscosity.

  To be more concrete, let us solve eq. (\ref{wave equation}) directly. In the hydrodynamic limit, $\omega, \ell \ll 1$, we can write the scalar modes in the perturbative series by using the expansion parameters $\omega$, $\ell$ \footnote{Usually, in perturbation expansion $\omega$ is treated as the same order as $\ell^{2}$ \cite{hydrodynamics}. However, in our case, if we treat $\ell^{2}$ as the expansion parameter, we will find there is a term proportional to $\ell$ in the equation. So we have to treat $\ell$ as the expansion parameter, not $\ell^{2}$.}. Express $\phi_{\ell}(u)=\phi_{0}(u)+\omega F_{1}(u)+\ell G_{1}(u)$, and keep to the order $O(\omega)$ and $O(\ell)$, we find the solutions for these three functions are all in the form of $C_{2}+\frac{1}{2}C_{1}\ln(-1+u^{2})$, where $C_{1}$ and $C_{2}$ are integration constants. Impose the boundary conditions, $\phi_{\ell}(0)=1$, and $\phi_{\ell}(u)$ is regular at the horizon $u=1$, we will find $\phi_{0}(u)=1$, and $F_{1}(u)=G_{1}(u)=0$. So to this order, the retarded Green's function $G^{R}=0$ and according to the Kubo's formula, the shear viscosity $\eta=0$. But this estimation is too naive since there are terms proportional to $\ell\omega$, and $\ell^{2}$, so this might imply that we should expand the equation to the second order. To the second order $\phi_{\ell}(u)=1+\omega F_{1}(u)+\ell G_{1}(u)+\ell \omega H(u)+\omega^{2}F_{2}(u)+\ell^{2}G_{2}(u)$, it is straightforward to see there is no imaginary part in the retarded Green's function and no contribution to the shear viscosity.  On the other hand, one can fix the null-like KK momentum $\ell$ fixed and finite so that we do not treat it as an expansion parameter. However, it is easy to see the retarded Green's function is also real, and the shear viscosity is zero.

   The seeming violation of the KSS bound $\eta/s \ge 1/4\pi$ in our discussions is deeply related to the singular horizon and our artificial regularity constraint of the scalar wavefunction on the horizon. Since the horizon is highly singular, it seems not possible for the wavefunction to comply with such a constraint. It is then interesting to see what is $\eta/s$ if we relax such a constraint in our case. For simplicity, we will just neglect the $\ell^2$ term in \eq{wave equation}, and obtain the solution
\be
\phi_\ell(u)=1+(C_2-i C_1 w) \ln(1-u^2), \qquad w:={\omega \ell R_{ads} \over 4\pi T}.
\ee
If we impose the regularity condition at the horizon, then $C_1=0$ and $\phi_{\ell}$ becomes real. Now we relax the condition and evaluate the flux, then we find
\be
{\eta\over s}={1\over 4\pi} {C_1 \ell R_{ads} \over 16\pi^2}.
\ee
This indeed indicates the correlation between singular horizon and the possible violation of the universal bound. Since we have no other reasonable boundary condition to fix $C_1$, we cannot say anything about the bound on $\eta/s$. Hopefully, some quantum gravity effect will resolve the horizon and fix $C_1$ in a subtle way.

\section{Conclusions and Discussions}
We propose a framework for non-relativistic holography. In our framework, the gravity dual to the co-dimension one non-relativistic CFT is the Newton-Cartan gravity with a negative cosmological constant. Using Bargmann lift to one-dimensional higher space-time, we construct the co-dimension two non-relativistic holography with covariant Einstein gravity. Conversely, one can also reduce from the Bargmann formulation to the Newton-Cartan one by performing the null-like KK reduction. The advantage of this formulation is that we can construct the same background metric with the Schr\"{o}dinger symmetry as obtained in \cite{Son:2008ye, Balasubramanian:2008dm} from the action principle.

We also construct the black hole solution with a null-like Killing vector in the bulk. The advantage to preserve the null-like Killing vector in the bulk is that the Galilean symmetry will be preserved under the holographic RG flow. However, due to the no-go theorem discussed in \cite{Liu:2003cta}, our black hole solution has a singular horizon. Although our black hole has a singular horizon \footnote{Other examples in using naked curvature singular solution for holography can be found in \cite{singular}.}, we can still have a well-defined temperature and thermodynamics. Therefore we incorporate this singular black hole in our Bargmann lift framework and derive from it the thermodynamics of the dual CFT. Besides, we have a correct thermodynamical behavior. Our temperature dependence of energy and entropy density match with the non-relativistic CFT. And our equation of state also matchs with dual non-relativistic CFT (at least for $d=4$). Besides, the particle number density is zero in our background, it means there is no particle condensation in our background. It would be interesting to see how to introduce the non-trivial vacuum of non-zero particle number density by turning on some bulk configuration.

Once we have a finite temperature background, one may wonder that if we can extract some transport coefficients such as the shear viscosity of the dual CFT in our background. Follow the recipe in \cite{Policastro:2001yc, Kovtun:2004de, hydrodynamics}, if we impose the regularity condition at the horizon we will get zero shear viscosity which violates the KSS bound $\frac{\eta}{s}\geq\frac{1}{4\pi}$. This is due to the singular horizon of our black hole, and actually if we relax the regularity condition at the horizon, we will obtain a finite shear viscosity. On the other hand, in \cite{Chen:2008ad} the authors calculate the viscosity for the cold atom system by resumming the fermion scattering diagrams in effective field theory approach, and they find the viscosity diverges except for $3+1$-dimensions.  Their result implies that the cold atoms not in $3+1$ dimensions cannot be a strongly coupled system. Intuitively, the formation of the threshold bound states of cold atoms at Feshbach resonance is equivalent to the divergent  scattering length, which implies strong correlation so that the viscosity should be vanishing. If this intuition is correct, then holographically it can be understood as caused by the singular horizon. However, it is in conflict with the results in \cite{Chen:2008ad} for the cold atoms. Therefore, we think the understanding of this issue in the context of non-relativistic holography remains open. Besides it is argued that the KSS bound can be violated in the large species of non-relativistic gas \cite{KSS bound violation}, it is interesting to realize this in the non-relativistic holography.

The generalization to the non-relativistic non-conformal theory should be starting with the usual near horizon geometry of the black p-brane with $p\ne 3$. Technically, the construction of the dual background geometry should be close to what have done in \cite{Lin:2006ie}, however, one should also impose the constraint to ensure the null-like Killing direction.

There are many other interesting issues worth further investigations. For example, it is interesting to introduce the finite chemical potential and finite density by turning on some bulk gauge and higgs fields. It is also interesting to consider the gravity dual to the non-relativistic theory for a matter field interacting with a Chern-Simons gauge field \cite{non-relativistic Chern-Simons}.

%%%%%%%%%%%%%%%%%%%%%%%%%%%%%%%%%%%%%%%%%%%%%%%%%%%%%
\section*{Acknowledgements}
%%%%%%%%%%%%%%%%

  We would like to thank Chi-Wei Wang who participated the project at its early stage. We thank Jiunn-Wei Chen, Veronika Hubeny, Wei Li, Shin Nakamura, Mukund Rangamani, Sang-Jin Sin, Ta-Sheng Tai, Wen-Yu Wen, Piljin Yi and K. P. Yogendran for helpful comments and discussions.  FLL also likes to thank for the hospitality of KIAS and IPMU where part of the work is completed. This work was supported by Taiwan's NSC grant 96-2112-M-003-014 and by NCTS.

%%%%%%%%%%%%%%%%%%%%%%%%%%%%%%%%%%%%%%%%%%%%%%%%%%%%%%%%

\end{document}